\documentclass[a4paper,12pt]{article}
\usepackage{authblk}
\usepackage{amsmath,amsfonts,amssymb}
\usepackage{graphicx}
\usepackage{geometry}
\usepackage{cite}
\usepackage{hyperref}
\usepackage[acronym]{glossaries}
\makeglossaries
\usepackage{enumitem}

\title{Lensless XUV Imaging Using Ghost Imaging on a Tabletop Platform}

\author[1,2,3]{Sukyoon Oh\thanks{Corresponding author: \href{mailto:sukyoon.oh@uni-jena.de}{sukyoon.oh@uni-jena.de}}}
\author[1,3]{Monalisa Mallick}
\author[4]{Thomas Siefke}
\author[1,2,3]{Tong Tian}
\author[1,2,3]{Christian Spielmann}
\affil[1]{Institute of Optics and Quantum Electronics, Abbe Center of Photonics, Friedrich Schiller University, Jena, Germany}
\affil[2]{GSI Helmholtz Centre for Heavy Ion Research, Darmstadt, Germany}
\affil[3]{Helmholtz Institute Jena, Fröbelstieg 3, 07743 Jena, Germany}
\affil[4]{Institute of Applied Physics, Abbe Center of Photonics, Friedrich Schiller University Jena, Germany}
\date{}

\newacronym{na}{NA}{Numerical Aperture}
\newacronym{gi}{GI}{Ghost Imaging}
\newacronym{hhg}{HHG}{High Harmonic Generation}
\newacronym{xuv}{XUV}{Extreme Ultraviolet}
\newacronym{sxr}{SXR}{Soft X-Ray}
\newacronym{cdi}{CDI}{Coherent Diffraction Imaging}
\newacronym{cgi}{CGI}{Computational Ghost Imaging}
\newacronym{mcgi}{MCGI}{Microscopic Computational Ghost Imaging}
\newacronym{tgi}{TGI}{Traditional Ghost Imaging}
\newacronym{cs}{CS}{Compressive Sensing}
\newacronym{cnr}{CNR}{Contrast-to-Noise Ratio}
\newacronym{fwhm}{FWHM}{Full Width Half Maximum}
\newacronym{slm}{SLM}{Spatial Light Modulator}
\newacronym{snr}{SNR}{Signal-to-Noise Ratio}
\newacronym{psnr}{PSNR}{Peak Signal-to-Noise Ratio}
\newacronym{dmd}{DMD}{digital micromirror device}
\newacronym{sgi}{SGI}{Sequential Ghost Imaging}
\newacronym{ds}{DS}{Different Step}
\newacronym{dd}{DD}{Different Directions}
\newacronym{dg}{DG}{Different Group}
\newacronym{xuvgi}{XUVGI}{Extreme Ultraviolet Ghost Imaging}
\newacronym{mse}{MSE}{Mean Squared Error}
\newacronym{n2v}{N2V}{Noise2Void}
\newacronym{n2vgi}{N2VGI}{Noise2Void-based Ghost Imaging}
\newacronym{ir}{IR}{Infrared}
\newacronym{fov}{FOV}{Field of View}

\begin{document}
\maketitle


\begin{abstract}
\noindent Correlation-based ghost imaging (GI) has attracted attention as a candidate framework for lensless imaging. However, its physical feasibility within tabletop extreme ultraviolet (XUV) environments remains insufficiently explored. Here, we present a lensless XUV microscopy platform integrating a compact tabletop high-harmonic generation (HHG) source with correlation-based ghost imaging, eliminating the need for imaging optics while evaluating the influence of mechanical and beam instabilities. By applying optimized post-processing techniques, we demonstrate image reconstruction of microscopic features under photon-limited conditions using a limited number of illumination patterns. This approach reduces system complexity and radiation dose, providing a baseline for accessible, low-dose XUV imaging in laboratory environments.
\end{abstract}

\section{Introduction}

\noindent \gls{xuv} imaging, operating in the sub-keV photon energy range, offers significant resolution improvements over visible-light microscopy owing to its shorter wavelengths, as described by the Abbe criterion \cite{Abbe1873}. This capability makes XUV imaging increasingly significant in materials science, life sciences, and nanotechnology. However, accessing these wavelength regimes often introduces substantial technical and financial challenges, since the mirrors, lenses, and gratings required for XUV imaging optics are scarce, lossy, and expensive \cite{Michette1993,Spielmann1997,Chapman2006,Thibault2008,attwood2000soft}.

\noindent Lensless imaging techniques circumvent this optics bottleneck by replacing imaging optics with computational reconstruction, and have become powerful alternatives for XUV microscopy. \gls{cdi} \cite{Miao2003,Zuerch2014,rothhardt2018table} records the far-field diffraction pattern of an isolated object and retrieves its real-space image through iterative phase-retrieval algorithms; as this inversion is ill-posed, CDI requires oversampled, highly coherent, high-flux illumination and strict spatial isolation of the sample to satisfy the finite-support constraint. Ptychography \cite{Chapman2010,holler2017high,loetgering2022advances} extends this concept by scanning a series of overlapping illumination spots across an extended sample, trading additional mechanical scanning precision and acquisition time for relaxed support constraints and quantitative phase retrieval over a wide field of view. Building on these principles, recent advances in XUV ptychography have achieved spatial resolutions below 34~nm by combining structured illumination with multiwavelength reconstruction algorithms using broadband \gls{hhg} sources \cite{eschen2024structured,pelekanidis2024illumination}, and have further been extended to biological specimens with resolutions down to 60~nm \cite{Liu2023}. However, both CDI and ptychography inherently demand high photon flux together with strict temporal and spatial coherence from the illumination source, and ptychography additionally requires precise, repeatable mechanical scanning, requirements that are difficult to satisfy simultaneously with compact, photon-limited tabletop HHG sources.

\noindent \gls{gi} offers a conceptually distinct approach that relaxes several of these requirements. In \gls{gi}, the object is sequentially illuminated with a set of known spatial patterns, and the total transmitted or reflected intensity for each pattern is recorded by a single-pixel (bucket) detector. Spatial information is recovered by correlating the measured bucket signals with the corresponding illumination patterns, rather than by directly recording a spatially resolved image at the detector \cite{Pittman1995,Abouraddy2001,Bennink2002,Cheng2004}. This approach allows image reconstruction without requiring a spatially resolving detector and provides an alternative to conventional imaging techniques.

\noindent \gls{gi}, a correlation-based form of single-pixel imaging (SPI), has since been extended to shorter wavelengths. In the ultraviolet range, computational and dark-field GI \cite{song2024computational}, photon-counting SPI \cite{ye2023ultraviolet}, and Hadamard SPI \cite{li2025hadamard} have been demonstrated, all with illumination patterns generated by a digital micromirror device (DMD). In the X-ray range, hard-X-ray GI with optimized transmission masks and learned reconstruction algorithms, as well as tabletop X-ray GI of moving objects with fast-spinning coding disks, have been reported \cite{he2020high,zeng2026tabletop}. These approaches cannot be transferred directly to the XUV range. To our knowledge, no commercial DMD operates in the XUV range. X-ray coding masks rely on thick absorbers on substrates that are transparent to X-rays, which would block XUV light, and their fabrication differs considerably from that of XUV masks. XUV masks require freestanding ultrathin membranes with a thickness of tens of nanometers, and experiments must be performed in a vacuum.

\noindent GI has also been demonstrated in the XUV regime using free-electron laser (FEL) sources \cite{Kim2020}. However, its implementation with a laboratory-scale tabletop \gls{hhg} source remains largely unexplored due to the lower photon flux compared with FEL-based systems. To address this gap, in this study we investigated the feasibility of lensless XUV ghost imaging by integrating a tabletop HHG source \cite{LHuillier1993,Brabec2000} with a single-pixel detection framework. We examined the main experimental limitations, particularly how the limited number of available illumination patterns and the spatial frequency content of the physical masks constrained spatial resolution and reconstruction quality. We also evaluated computational post-processing strategies to mitigate these limitations and established a quantitative baseline for future laboratory-scale XUV-GI systems. Together, these results assessed the practical feasibility of tabletop XUV-GI and identified the key factors limiting its performance under realistic experimental conditions.

\section{Materials and Methods}
\subsection{Experimental Setup}

\begin{figure}[h!]
\centering
\includegraphics[width=1\linewidth]{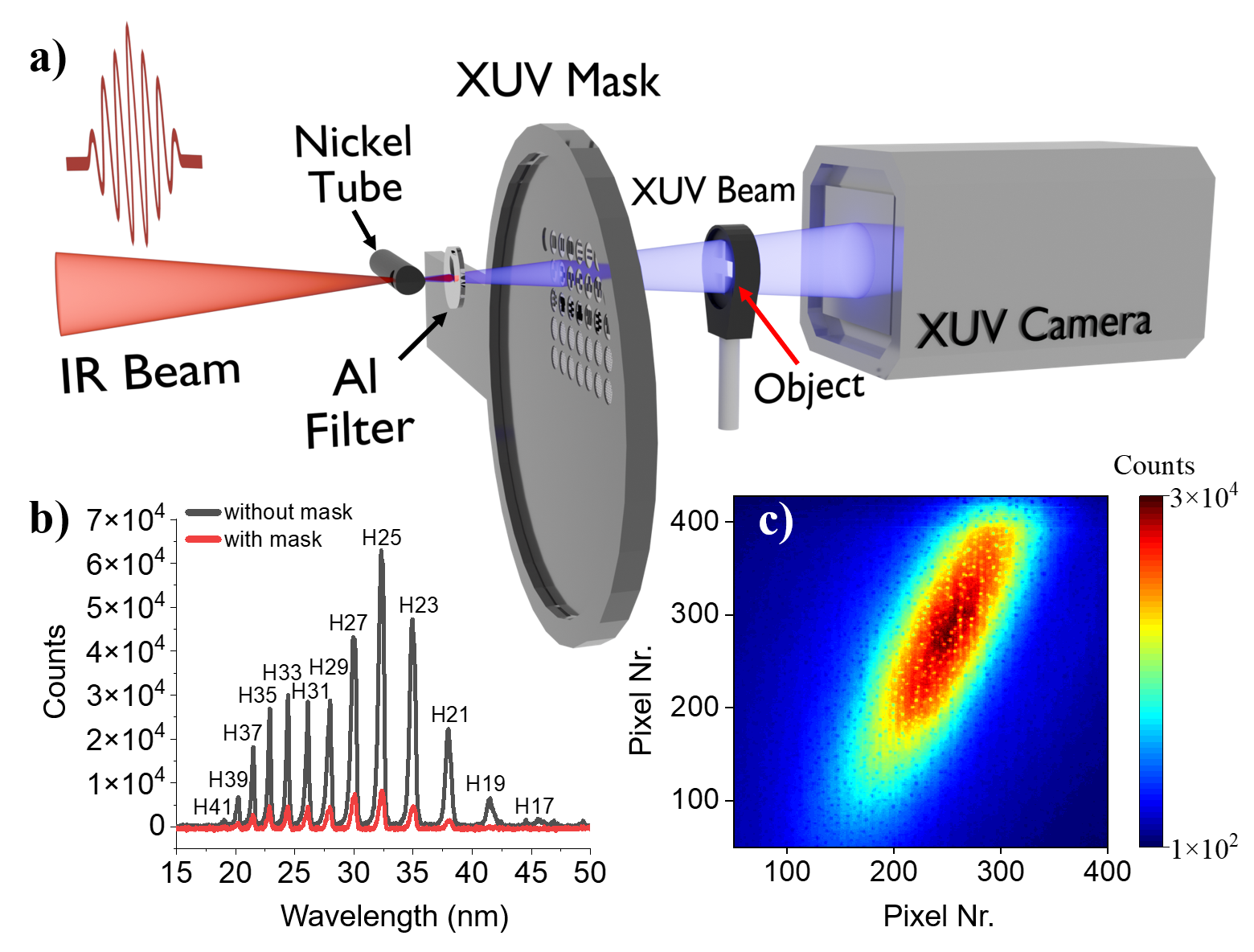}
\caption{\textbf{XUV ghost-imaging experimental setup and illumination characteristics.}
(a) Schematic of the experimental setup. A broadband XUV beam generated by high-harmonic generation (HHG) in an argon-filled nickel tube is spectrally filtered using a 500~nm-thick aluminum (Al) filter and transmitted through the XUV transmission mask and target object before detection.
(b) Measured XUV emission spectra. The labeled peaks (H17--H41) correspond to odd-order harmonics of the 800~nm driving field. The black curve was recorded without the mask, whereas the red curve was recorded with the mask positioned at an unpatterned region containing a 2-mm-diameter opening.
(c) Spatial profile of the XUV beam recorded by the XUV camera without the mask in the beam path.}
\label{fig_exp_setup}
\end{figure}

\noindent Figure~\ref{fig_exp_setup}a shows the layout of the tabletop XUV ghost-imaging setup. A broadband XUV beam is sequentially transmitted through a structured XUV mask and the target object before being recorded by an XUV-sensitive, pixel-array detector. No image-forming optics, such as mirrors, lenses, or zone plates, are placed between the mask, the object, and the detector. The complete tabletop XUV ghost imaging configuration is shown in Fig.~\ref{fig_exp_setup}a, with the measured broadband XUV spectrum and spatial beam profile depicted in Figs.~\ref{fig_exp_setup}b and \ref{fig_exp_setup}c. The broadband XUV beam sequentially passes through the structured XUV mask and the target object before reaching the XUV detector. The XUV mask is mounted on a high-precision motorized linear XY-stage to project spatially structured illumination patterns. The target image is subsequently reconstructed by computing correlations between the translated mask positions and the integrated intensity fluctuations recorded by the detector. This scheme is lensless because spatial information is not recovered by focusing light onto the detector, but is instead reconstructed computationally from the second-order correlation between the structured illumination patterns and the corresponding integrated bucket intensity recorded for each pattern. In this sense, the detector does not need to resolve the object spatially; it is operated as a single-pixel bucket detector by integrating the recorded signal over the illuminated region of interest for each exposure, even though the physical sensor used here is a pixel-array XUV camera (an in-vacuum, Peltier-cooled scientific CCD from the greateyes GE-VAC~1024$\times$1024 series; $1024\times1024$ pixels, $13\times13~\mu\text{m}^2$ pitch). In conventional computational GI, the patterns are displayed on a spatial light modulator or DMD and the bucket signal is recorded by a photodiode. Such programmable modulators are not available in the XUV range, so static transmission masks translated by an XY stage are used instead. Integrating the camera pixels over the illuminated region is mathematically equivalent to a single-pixel measurement, and a dedicated XUV photodiode could replace it in future implementations. The camera additionally provides the measured illumination patterns and the beam footprint used in Section~3.2. 

\noindent We employ a compact tabletop HHG source driven by nonlinear frequency conversion in noble gas media \cite{McPherson1987}. The source utilizes a Ti:Sapphire laser system operating at a center wavelength of 800~nm (1~kHz repetition rate, 0.5~mJ pulse energy, 35~fs pulse duration) focused into an argon-filled nickel tube at a backing pressure of 70~mbar. A self-drilled entrance aperture maintains stable phase-matching conditions throughout operation.

\noindent At the focal point, the driving infrared beam diameter is approximately 40~$\mu$m, corresponding to a peak intensity of $2 \times 10^{15}~\text{W/cm}^2$. Residual infrared radiation is effectively blocked by a 500-nm-thick aluminum filter. The resulting broadband XUV emission delivers a total photon flux of $1.2 \times 10^{9}$~photons/s before entering the XUV mask, which is lower than typical synchrotron-based microscopy sources \cite{attwood2000soft}.

\noindent To assess the practical viability of XUV-GI under such photon-limited conditions, we evaluated the minimum photon budget required for image formation. The minimum photon counts required to achieve spatial resolution at a $\text{SNR} \ge 5$ were estimated using the Rose criterion \cite{rose1975vision}. For an $80 \times 80$ pixel grid under equivalent transmission and detector efficiency assumptions, the calculated minimum required photon counts were $1.12 \times 10^{7}$ for XUV-GI, $2.75 \times 10^{10}$ for HHG-based CDI \cite{dinh2015coherent}, and $3.51 \times 10^{14}$ for synchrotron-based soft X-ray scanning transmission microscopy (STXM) \cite{spath2015confocal}. Detailed parameter settings and step-by-step calculations are provided in Supplementary Note 1.

\noindent Figure~\ref{fig_exp_setup}b compares the XUV spectrum measured without the mask (black) to that measured through an unpatterned, 2-mm-diameter opening in the mask substrate (red). Both spectra were recorded with an XUV spectrometer using an Andor Newton CCD camera ($2048\times512$ pixels, $13.5\,\mu\text{m}$ pixel size). This comparison shows the flux that passes through the mask substrate before the Hadamard patterns are applied. The red spectrum is lower because the XUV beam is larger than the opening, which therefore transmits only the part of the beam within its 2-mm diameter. The labels H17 to H41 denote the odd harmonics of the 800~nm driving field. The variation of the counts with wavelength mainly reflects the harmonic emission spectrum, the wavelength-dependent transmission of the aluminum (Al) filter, and the spectral response of the spectrometer and its detector.

\noindent Figure~\ref{fig_exp_setup}c shows the spatial beam profile recorded by the camera without the mask in the beam path. Its elliptical profile arises from the non-ideal spatial intensity distribution of the driving Ti:Sa laser beam, its focusing geometry, and the position relative to the nickel-tube gas cell, which was adjusted to expand the XUV footprint over the $1\times1~\text{mm}^2$ mask area. Since the patterns sample only the central region, the outer ellipticity has a minor effect on reconstruction, with remaining non-uniformities addressed by histogram equalization (Section~3.2). The small dark dots at the center correspond to the most intense region of the beam, while other spots across the profile may arise from defects or dust contamination on the Al filter or detector surface.

\subsection{Ghost Imaging Reconstruction Algorithm and Patterns}

\noindent The \textbf{\gls{tgi}} reconstruction algorithm utilizes the second-order correlation between two recorded datasets to reconstruct an object's image. In this framework, total intensity values collected by a single-pixel detector are correlated with known 2D intensity distributions of the illumination patterns. The reconstruction formula is expressed as:

\begin{equation}
G(x,y) = \frac{1}{N} \sum_{n=1}^{N} [I_n(x,y) - \langle I \rangle][S_n - \langle S \rangle]
\label{eq:tgi}
\end{equation}

\noindent where the reconstructed object image, $G(x,y)$, is obtained by computing the second-order correlation function \cite{oh2023improvements,oh2025efficient} between the spatial intensity distribution $I_n(x,y)$ projected onto the sample and the corresponding bucket intensity $S_n$ recorded by the XUV detector. Here, $N$ denotes the total number of illumination patterns evaluated in the experiment.\\

\noindent \textbf{\gls{cs}} enables robust image reconstruction from significantly fewer measurements than required by traditional sampling theorems \cite{Katz2009,Hahamovich2021}. In GI, the recorded signal vector $S$ from the bucket detector is formulated in matrix form as $S = I \times O$, where $I$ represents the flattened matrix of illumination patterns and $O$ represents the flattened unknown target image. To recover $O$, an optimization problem is solved by exploiting sparsity in a transform domain, typically formulated as an $\ell_{1}$-norm minimization problem:

\begin{equation}
\hat{O} = \arg\underset{O}{\min} \| O \|_{1} \quad \text{subject to} \quad I \times O = S
\label{eq:cs}
\end{equation}

\noindent where $\|O\|_{1}$ denotes the $\ell_{1}$-norm of $O$. The constraint $I \times O = S$ ensures strict consistency between the reconstructed image and the physically measured bucket signals.\\

\noindent Reconstruction via \textbf{Differential Hadamard Patterns.} Hadamard patterns offer deterministic, orthogonal spatial sampling, which is structurally superior to random speckle patterns. Differential Hadamard schemes capture the intensity difference between complementary positive and negative pattern sets, thereby enhancing contrast recovery \cite{Zhang2017}:

\begin{equation}
G(x,y) = \mathcal{H}^{-1}\left[\, h(u,v) = D_{+1} - D_{-1} \,\right]
\label{eq:had}
\end{equation}

\noindent A set of $N$ Hadamard patterns is generated, where each pattern is arranged on a $\sqrt{N}\times\sqrt{N}$ grid, so that $N$ defines the number of independent spatial modes and thus the resolution of the reconstruction. Each pattern is realized by a complementary mask pair, $D_{+1}$ and $D_{-1}$, and the differential pattern is obtained by subtracting the two, which cancels contributions common to both exposures, such as background light and slow drifts of the source intensity. We fabricated both mask sets to compare how the number of exposures, 64 patterns versus 128 (64 complementary pairs), affects the image quality. The recorded detector intensities are compiled into a signal vector corresponding to the differential patterns, and the spatial image $G(x,y)$ is recovered by applying an inverse Hadamard transform $\mathcal{H}^{-1}$ to the resulting differential signal. An alternative single-polarity scheme, which projects only the $N$ positive patterns and subtracts a suitably normalized background or reference signal, can thus recover equivalent object information with approximately half the exposures \cite{yu2021differential,moodley2023all}.

\subsection{XUV Mask Design and Fabrication}

\begin{figure}[h!]
\centering
\includegraphics[width=0.9\linewidth]{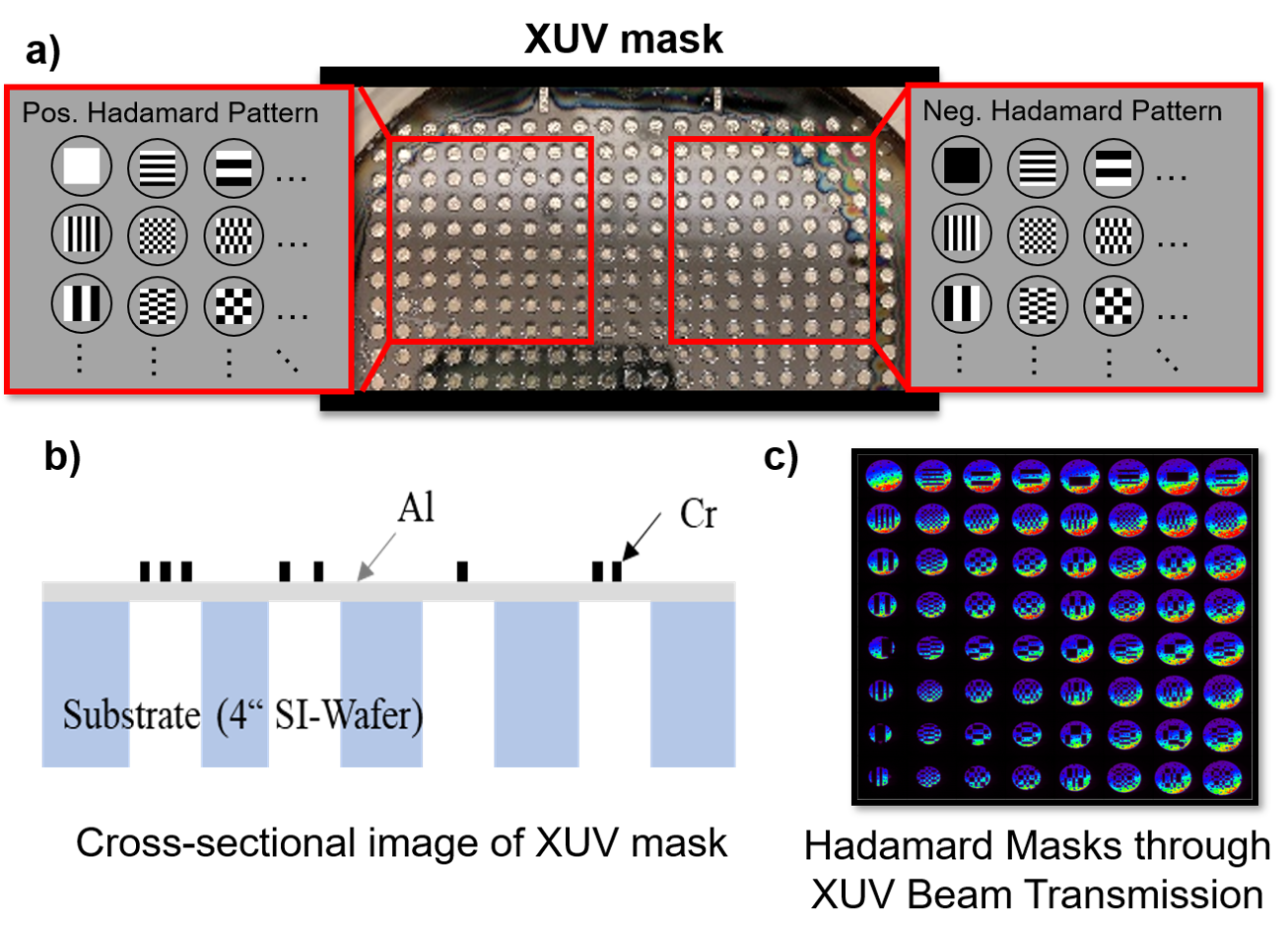}
\caption{\textbf{Fabricated XUV Micro-Mask Structure.} (a) Photograph of the fabricated mask with Hadamard patterns. (b) Cross-sectional diagram illustrating the mask layer geometry. (c) XUV transmission profile recorded across different mask regions.}
\label{fig_mask}
\end{figure}

\noindent For the differential scheme described above, complementary sets of $N=64$ positive and negative Hadamard patterns were fabricated as XUV transmission masks. Optimal mask performance requires high attenuation contrast between transparent and opaque regions. We utilized a 100~nm thick aluminum membrane substrate ($\sim 80\%$ transmission across the 20--60~nm spectral range) coated with a 100~nm thick chromium (Cr) layer to suppress unwanted background transmission \cite{Henke1993}.

\noindent Because fabricating a new photomask would have required prohibitive cost and time, these pattern sets were realized on an existing photomask layout. This fixed layout, together with the limited travel range of the vacuum stage, restricted the number of distinct patterns accessible in a single measurement run to $N=64$. The layout was defined on a 4-inch silicon wafer substrate, which was partitioned into functional regions with different feature sizes, and the Hadamard sets used here have a minimum feature size of 125~$\mu$m. Each pattern covers an area of $1 \times 1~\text{mm}^2$ within a 2~mm diameter aperture, and this area determines the \gls{fov}.

\noindent Micro-fabrication of the mask array (Institute of Applied Physics, Jena) divided the silicon wafer into these dedicated functional zones (Fig.~\ref{fig_mask}a). Figure~\ref{fig_mask}b shows a cross-sectional schematic of the material layer structure. The processing workflow concluded with photoresist removal and deep backside silicon etching. Fabricated mask quality and attenuation performance were confirmed under direct XUV beam exposure (Fig.~\ref{fig_mask}c).

\noindent To confirm that diffractive spreading does not distort the projected illumination patterns over the propagation distance, Fresnel numbers were calculated to evaluate optical propagation regimes across the detector plane (Supplementary Note 2) \cite{Gu2000}. As illustrated in Supplementary Fig.~S1, the object was positioned 2.5~cm from the CCD sensor, operating the Hadamard illumination masks well within the near-field (Fresnel) diffraction regime.

\noindent With the fabricated masks, the 64 patterns form a complete Hadamard basis of order 64, which corresponds to an $8\times8$ grid of spatial modes. With the $1\times1~\text{mm}^2$ pattern area, each element of this grid corresponds to $125~\mu\text{m}$. How this number translates into image resolution depends on the reconstruction method.

\noindent For the differential reconstruction using the inverse Hadamard transform in Eq.~\eqref{eq:had} and the \gls{cs} reconstruction in Eq.~\eqref{eq:cs}, the image is reconstructed directly in this 64-mode basis. The resolution is therefore limited by the number of patterns, where 64 patterns yield an $8\times8$ image regardless of the camera's pixel resolution.

\noindent For the \gls{tgi} reconstruction in Eq.~\eqref{eq:tgi}, in contrast, the image is obtained by correlating the bucket signal with the pattern images $I_n(x,y)$. The output resolution is thus set by the pixel grid of these pattern images rather than by the number of patterns. When the ideal binary mask design is used for $I_n(x,y)$, the resulting image has an $8\times8$ pixel grid. If the patterns recorded by the XUV camera are used instead (Section~3.2, Measured illumination pattern), $I_n(x,y)$ is defined on the camera's native $80\times80$-pixel grid, and the TGI image is accordingly reconstructed on the same $80\times80$ grid. This explains why the reconstructions in Figs.~\ref{fig_post_process} and \ref{fig_final_recon} are displayed at $80\times80$ pixels.

\noindent However, the $80\times80$ reconstruction grid does not represent 6,400 independently resolved spatial elements. A fully sampled reconstruction of an arbitrary image on this grid would require up to 6,400 independent illumination patterns, whereas only 64 patterns are available in our experiment. Thus, the measurement set covers only 1\% of the measurements required for full sampling, leaving most spatial degrees of freedom underdetermined without additional prior information or assumptions \cite{gibson2020single}. The $80\times80$ grid therefore represents the reconstruction grid rather than the number of independently resolved spatial elements.

\noindent Accordingly, increasing the number of fabricated Hadamard patterns (e.g., to 256 for a $16\times16$ grid) would proportionally increase the number of independently resolvable elements across all three reconstruction approaches. A previous X-ray ghost imaging study \cite{he2020high} demonstrated reconstructions of $32\times32$-pixel images using 1,024 patterns and $64\times64$-pixel images using 4,096 patterns, confirming that higher spatial resolution is directly achieved by scaling the pattern count.

\section{Results}
\subsection{XUV Ghost Imaging: Experimental Results}

\begin{figure}[h]
\centering
\includegraphics[width=1\linewidth]{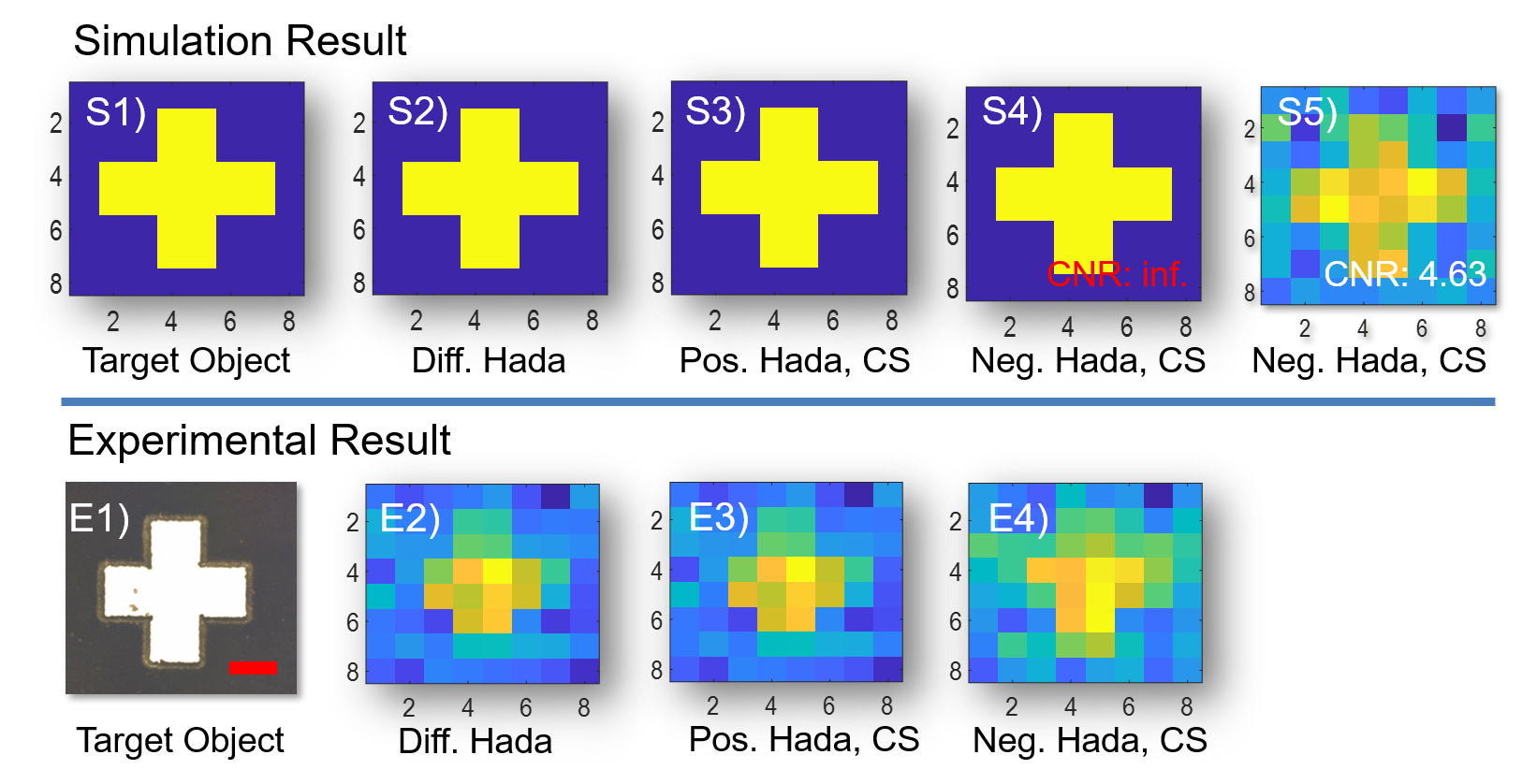} 
\caption{\textbf{Numerical simulations and experimental XUV--GI reconstructions of a cross target.} Top row: Numerical simulations. (S1) Target object used in the numerical simulation. (S2) Reconstructed image obtained using differential Hadamard patterns with an inverse Hadamard transform. (S3, S4) Reconstructed images using (S3) positive and (S4) negative Hadamard patterns combined with \gls{cs}. (S5) Reconstructed image obtained with the negative Hadamard patterns and \gls{cs} under a 1-pixel spatial misalignment between the simulated detector signal and the pattern used for reconstruction (\gls{cnr}: 4.63). Bottom row: Experimental ghost imaging reconstructions of a cross target. (E1) Optical microscope image of the cross target with a 200~$\mu$m scale bar. (E2) Experimental inverse Hadamard transform reconstruction using differential Hadamard patterns. (E3, E4) Experimental \gls{cs} reconstructions using (E3) positive and (E4) negative Hadamard patterns.}
\label{fig_ex_simu_gi}
\end{figure}

\noindent The primary objective of this study is to demonstrate the feasibility of tabletop XUV-GI under conditions of limited pattern count and spatial resolution. The top row of Figure~\ref{fig_ex_simu_gi} shows numerical simulation results for a cross-shaped target object (Fig.~\ref{fig_ex_simu_gi} S1). Differential Hadamard reconstruction using 64 complementary positive-negative pattern pairs (128 physical exposures) in Fig.~\ref{fig_ex_simu_gi} S2 accurately recovers the target geometry, and \gls{cs} reconstructions using either 64 positive or 64 negative patterns (Figs.~\ref{fig_ex_simu_gi} S3 and \ref{fig_ex_simu_gi} S4) recover it equally well. In both the simulation and the experiment, the CS reconstructions use 64 patterns, and the differential reconstruction uses the same 64 patterns together with their complements. The complements double the number of exposures to 128 but do not increase the number of independent patterns, so the spatial resolution is set by the 64 patterns in all cases. In the simulation, a spatial misalignment of even a single pixel between the patterns used for reconstruction and the simulated detector signal reduces the \gls{cnr} from infinity for the noise-free, perfectly aligned reconstruction (Fig.~\ref{fig_ex_simu_gi} S4) to 4.63 (Fig.~\ref{fig_ex_simu_gi} S5; further detailed analysis of spatial misalignment is presented in Section~3.2).

\noindent The bottom row of Fig.~\ref{fig_ex_simu_gi} shows the experimental reconstruction results obtained for a cross-shaped object with a 200~$\mu$m scale bar. Figure~\ref{fig_ex_simu_gi} E2 displays the result retrieved through the differential Hadamard reconstruction method, while the reconstructed images in Figs.~\ref{fig_ex_simu_gi} E3 and \ref{fig_ex_simu_gi} E4 were obtained using \gls{cs} combined with positive and negative Hadamard patterns, respectively. Although low pixel resolution and noticeable pixelation artifacts are present due to the limited number of illumination patterns, the fundamental spatial structure of the object is nevertheless recovered.

\noindent Interestingly, the reconstructed images in Figs.~\ref{fig_ex_simu_gi} E2 and \ref{fig_ex_simu_gi} E3 appear visually almost identical, although E2 uses all 128 exposures (64 complementary pattern pairs) and E3 only the 64 positive patterns. This result shows that simply increasing the number of exposures does not improve the image quality. It is expected because the negative Hadamard pattern is the exact complement of the positive one, so the negative-pattern signal can be derived from the positive-pattern signal and the open-field signal ($D_{-1} = D_{0} - D_{+1}$). The differential signal in Eq.~\eqref{eq:had} ($D_{+1} - D_{-1} = 2D_{+1} - D_{0}$) can therefore be obtained from the positive patterns alone~\cite{yu2021differential,moodley2023all}. Improving the image quality instead requires increasing the number of independent patterns in the measurement set.

\noindent Although these initial measurements verify basic target shape recovery, noticeable discrepancies remain when compared with ideal numerical simulations (Figs.~\ref{fig_ex_simu_gi} S2--S4). This image degradation primarily stems from experimental factors, including spatial misalignment between the patterns used for reconstruction and the actual signals recorded by the detector, the limited spatial frequency content of the Hadamard patterns, and subtle beam-pointing fluctuations inherent to the HHG source.

\noindent To overcome these physical constraints, we implemented a dedicated post-processing pipeline designed to realign and filter the raw experimental datasets. To verify the generalizability and spatial resolution limits of this framework across diverse sample features, extended evaluations were also performed. Supplementary Note 3 and S.Figs.~2--3 provide further evaluations on smaller target dimensions and alternative target geometries.

\subsection{Data Processing Pipeline}

\begin{figure}[h]
\centering
\includegraphics[width=1\linewidth]{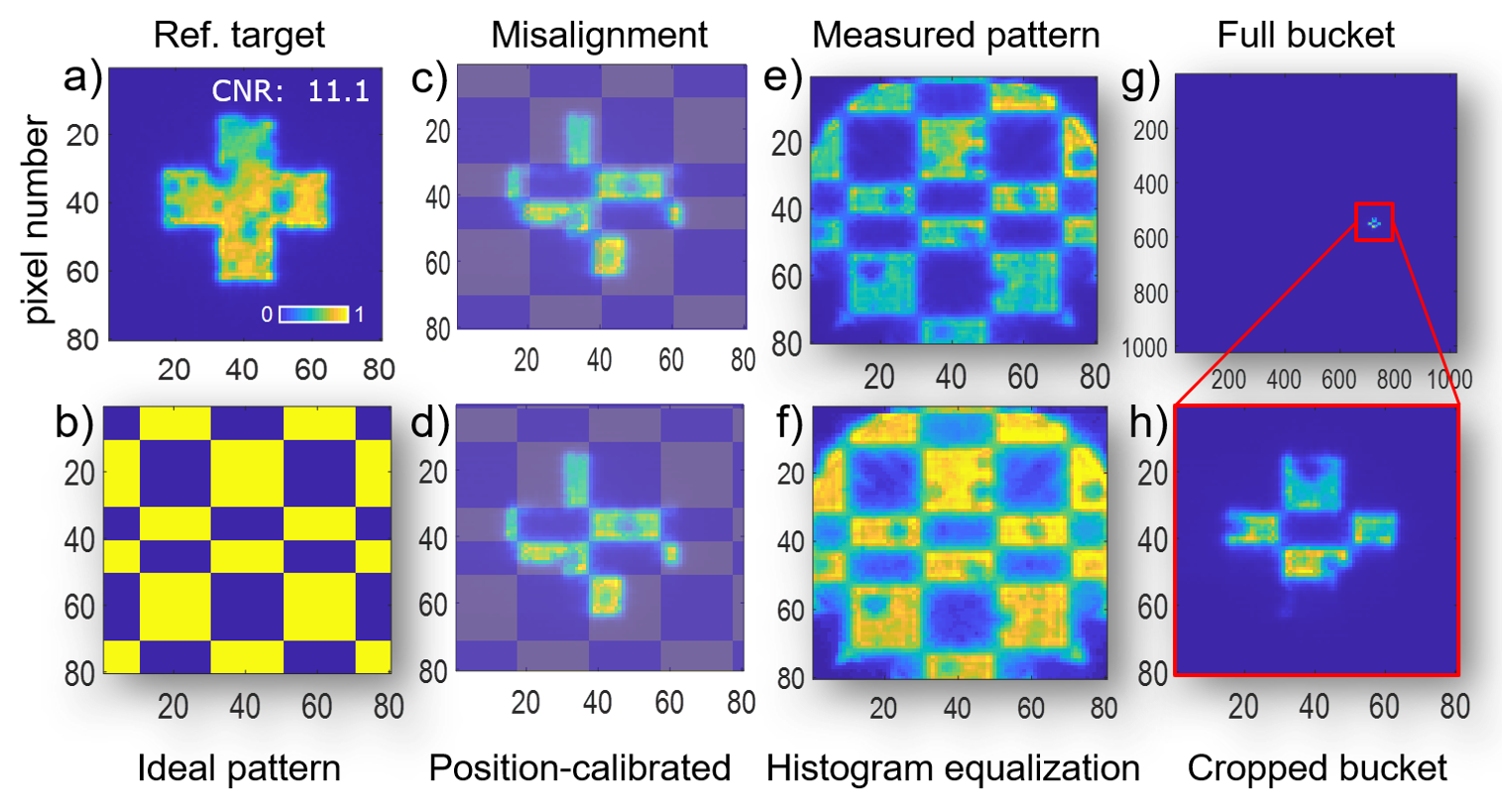}
\caption{\textbf{Post-processing pipeline for XUV ghost image reconstruction.} (a) Reference image obtained directly with the XUV camera. (b) Ideal numerical pattern. (c) Spatial misalignment between the ideal pattern and the signals recorded by the detector. (d) Pattern overlay after positional calibration. (e, f) Measured illumination pattern before (e) and after (f) histogram equalization. (g, h) Bucket integration domain over the full detector frame (g) and over the cropped active region (h).}
\label{fig_post_process}
\end{figure}

\noindent To minimize systematic reconstruction errors, a sequence of corrective post-processing steps was incorporated into the reconstruction workflow. In tabletop XUV-GI, spatial misalignment and beam-profile non-uniformities inevitably arise. Unlike a conventional single-pixel detector, the pixel-array detector (XUV camera) allows these effects to be visualized directly, which provides a reliable benchmark for evaluating the physical feasibility of the imaging process. A baseline reference image (Fig.~\ref{fig_post_process}a) was acquired without spatial modulation patterns. Its \gls{cnr} of 11.1 serves as a reference value for comparison with the reconstructed ghost images.

\medskip
\noindent \textbf{Positional calibration:} Mechanical tolerances in the translation stage and the XUV mask mounting assembly introduce lateral offsets between the ideal numerical patterns used for reconstruction and the corresponding signals recorded by the detector (Fig.~\ref{fig_post_process}c). To compensate for this mismatch, the reconstruction patterns were laterally shifted to the measured spatial positions (Fig.~\ref{fig_post_process}d), which reduces the reconstruction errors caused by mechanical misalignment. Because the mask position is set mechanically, residual offsets of this kind cannot be removed by alignment alone, and the numerical calibration plays the role of registering the measured pattern positions to the known pattern geometry.

\medskip
\noindent \textbf{Measured pattern:} Because \gls{gi} relies on second-order intensity correlations, replacing the ideal binary illumination patterns with the experimentally measured intensity distributions improves the correlation fidelity. Image reconstruction was therefore performed using the illumination patterns directly recorded by the XUV camera (Fig.~\ref{fig_post_process}e), which represent the actual illumination conditions more accurately.

\medskip
\noindent \textbf{Histogram equalization:} Shot-to-shot beam-pointing fluctuations of the \gls{hhg} source cause spatial non-uniformities and intensity variations among successive illumination patterns (Fig.~\ref{fig_post_process}e). To reduce these variations, histogram equalization was applied to normalize the intensity distribution within the illuminated beam footprint (Fig.~\ref{fig_post_process}f). This normalization suppresses reconstruction noise from beam-pointing jitter and spatial inhomogeneity and improves the pattern-to-pattern consistency.

\medskip
\noindent \textbf{Bucket domain selection:} Integrating the signal over the entire $1024 \times 1024$ detector array accumulates dark current and electronic readout noise from unilluminated sensor regions, which can obscure the target-modulated signal (Fig.~\ref{fig_post_process}g). Bucket integration was therefore restricted to the illuminated beam footprint (Fig.~\ref{fig_post_process}h), which isolates the target information and suppresses the integrated background noise. These results indicate that \gls{gi} requires neither a large detection area nor a high photon flux for reliable image reconstruction.

\medskip
\noindent Notably, a recognizable target reconstruction (Figs.~\ref{fig_ex_simu_gi} E2--E4) remains possible even without positional calibration or histogram equalization. This demonstrates that tabletop XUV-GI can recover basic target features under photon-limited and experimentally non-ideal conditions.

\subsection{Analysis of Reconstruction Constraints}
\label{sec:results_summary}

\begin{figure}[h]
\centering
\includegraphics[width=1\linewidth]{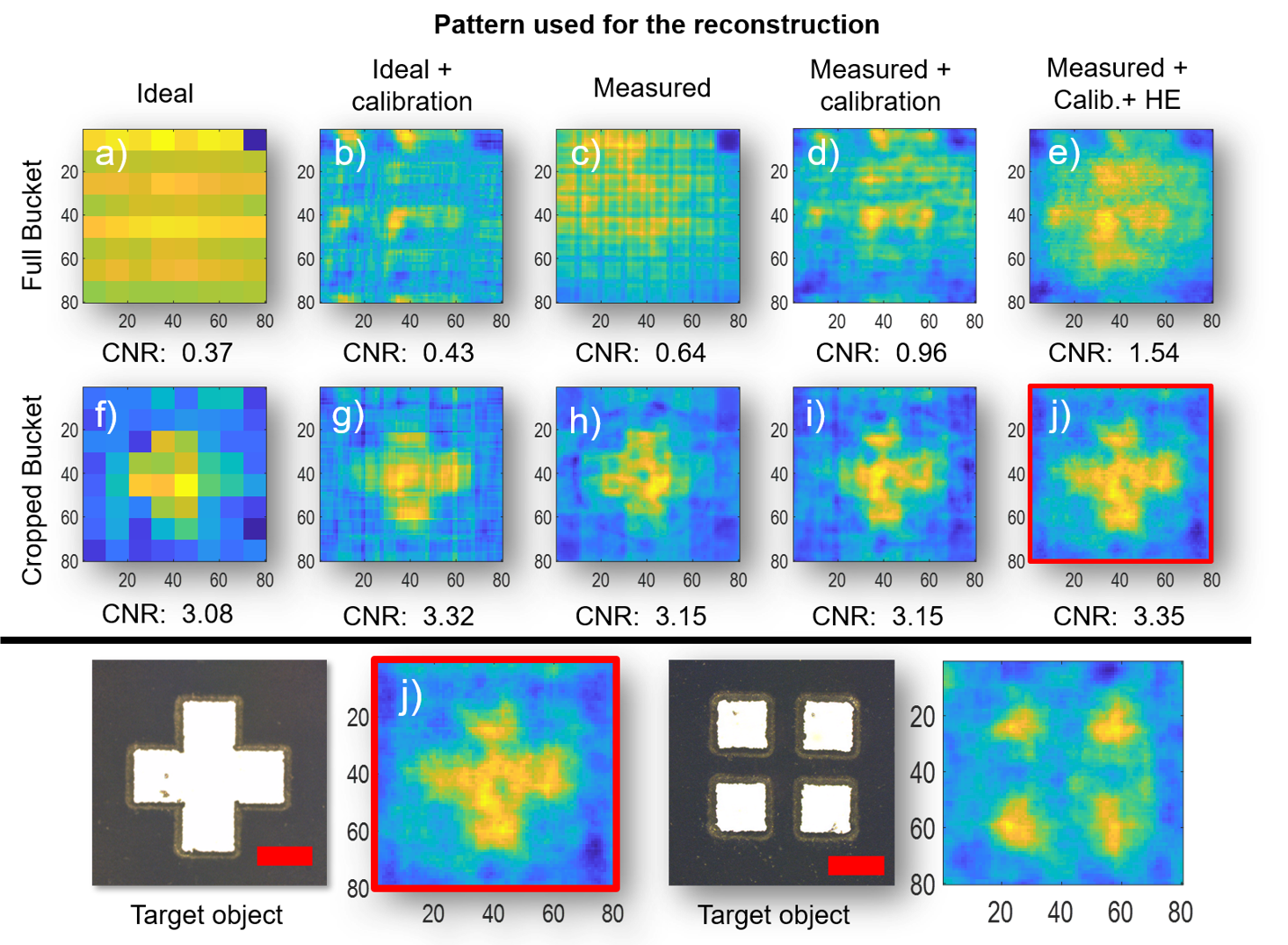}
\caption{\textbf{Comparison of processing methods and experimental ghost imaging reconstructions.} (a--e) Reconstructions using the full-frame bucket signal. (f--j) Reconstructions using the cropped bucket signal. In each row, the patterns used for the reconstruction are, from left to right: ideal pattern, position-calibrated ideal pattern, measured pattern, position-calibrated measured pattern, and position-calibrated measured pattern with histogram equalization (HE). The CNR is given below each panel. (Bottom row)~Optical microscope images of the targets (red scale bars: $200\,\mu\text{m}$) and the corresponding reconstructions with the best condition (j) for a single-cross target (left) and a four-square target array (right).}
\label{fig_final_recon}
\end{figure}

\noindent To quantify the effect of each post-processing step on image quality, reconstructions were performed for all processing conditions using the \gls{tgi} framework (Fig.~\ref{fig_final_recon}).

\noindent With the full-frame bucket signal, the target geometry is hardly visible under most conditions (Figs.~\ref{fig_final_recon}a--e). The \gls{cnr} increases from 0.37 for the ideal patterns (Fig.~\ref{fig_final_recon}a) to 1.54 for the position-calibrated measured illumination patterns with histogram equalization (Fig.~\ref{fig_final_recon}e).

\noindent Restricting the bucket integration domain to the active beam footprint increases the \gls{cnr} for all pattern conditions (Figs.~\ref{fig_final_recon}f--j). The cross-shaped geometry can be identified, and the maximum \gls{cnr} is 3.35 for the position-calibrated measured illumination patterns with histogram equalization (Fig.~\ref{fig_final_recon}j). Cropping alone, without any other processing step, already increases the \gls{cnr} obtained with the ideal patterns from 0.37 to 3.08 (Figs.~\ref{fig_final_recon}a and \ref{fig_final_recon}f).

\noindent As shown in the bottom row of Fig.~\ref{fig_final_recon}, the final reconstruction shows the main structure of the target, but its quality remains limited. Further results for four-square samples and smaller feature sizes are given in Supplementary Note 3.

\noindent Besides image quality, the required illumination exposure is important for assessing the feasibility of imaging radiation-sensitive specimens. Based on the target transmission and the total exposure time of 19.2~s, the radiation dose delivered to a biological model target (density: $1.35~\text{g/cm}^3$) was estimated to be $2.5 \times 10^3$~Gy. This value is well below the typical doses reported for EUV ptychography \cite{Liu2023} and several orders of magnitude below the established Henderson limit of $2 \times 10^7$~Gy for structural damage \cite{henderson1990cryo}.

\section{Discussion and Conclusion}

\noindent In this study, we evaluated the physical feasibility of an XUV ghost imaging system driven by a tabletop HHG source. Under current experimental conditions, the spatial resolution is primarily limited by the number of available patterns ($N=64$) and the physical feature size of the microfabricated patterns ($125~\mu\text{m}$). Within these operational constraints, the system achieved a baseline \gls{cnr} of 3.35, establishing a quantitative reference for future optimization.

\noindent While post-processing strategies, such as positional calibration and restricting the signal to the illuminated region of the detector, effectively mitigate noise, they cannot overcome the underlying physical sampling limits. Consequently, achieving higher image fidelity requires nanometer-scale mask arrays with finer features to expand the spatial bandwidth, alongside a substantial increase in the pattern count. These experimental outcomes collectively provide a solid foundation for designing advanced laboratory-scale XUV ghost imaging architectures.

\noindent A central advantage of XUV imaging is that its short wavelength theoretically permits a spatial resolution approaching the nanometer scale. In practice, however, the resolution is currently set by the number of patterns and the size of the smallest mask element rather than by the wavelength. For example, to reconstruct an image of $320\times200$ pixels with a pixel size of $30\times30~\text{nm}^2$ (comparable to the XUV wavelength), the mask set must satisfy two conditions. First, it must contain approximately $64{,}000$ patterns, one for each pixel, which is three orders of magnitude more than the current 64. Second, each pattern must consist of $30\times30~\text{nm}^2$ elements arranged over an area of $9.6\times6~\mu\text{m}^2$. The primary bottleneck is therefore the fabrication of such a large set of fine XUV masks.

\noindent Finally, while the present scheme is sensitive solely to the object's amplitude transmission, phase-sensitive ghost imaging could in principle be extended to the XUV regime. This expansion relies on the development of suitable XUV phase masks combined with the high spatial coherence of HHG sources \cite{Zhang2017,sephton2023revealing}. Assessing the practical feasibility of fabricating such phase masks under current nanofabrication constraints remains an important direction for future work.

\section*{Supplementary Material}
See the supplementary material for detailed experimental parameters and additional dataset analysis.

\section*{Acknowledgements}
We thank Chang Liu for helpful discussions. Sample fabrication was carried out by the microstructure technology team at IAP Jena, whose support and facilities are gratefully acknowledged. A Large Language Model was used for language and grammar refinement. All authors have read and approved the final manuscript.

\section*{Author Contributions}
S.O. performed conceptualization, methodology, investigation, formal analysis, software, writing of the original draft, and visualization; M.M. contributed to investigation and formal analysis; T.S. provided resources; T.T. provided software and review; and C.S. handled formal analysis, validation, funding acquisition, writing -- review \& editing, and project supervision.

\section*{Funding}
Deutsche Forschungsgemeinschaft (DFG, German Research Foundation) under Germany’s Excellence Strategy—EXC 2051 (Project-ID 390713860), ``Balance of the Microverse''; The Free State of Thuringia within the project ``Quantum Hub Thüringen'' (Funding ID 2021 FGI 0044); DAAD (Deutscher Akademischer Austauschdienst) German Academic Exchange Service, Funding programme/-ID: (57552340) Research Grants---Doctoral Programmes in Germany, 2021/22.

\section*{Availability of Data and Materials}
Data available on request due to restriction of privacy. The data presented in this study are available on request from the corresponding author. The data is not publicly available due to related experiments still in progress and involve unpublished papers.


\section*{Declarations}

\textbf{Competing interests:} The authors declare that they have no known competing financial interests.

\bibliographystyle{unsrt}
\bibliography{references}

\end{document}